# Adaptive Fault Diagnosis using Self-Referential Reasoning

by Robert Cowen*

1. Introduction.

I first encountered logical puzzles about "Knights," who always tell the truth, "Knaves," who always lie, and "Normals," who sometimes tell the truth and sometimes lie, in Raymond Smullyan's wonderful book [7], "What is the Name of this Book?" In the book, Ray not only entertains, but takes the reader on a journey, visiting many bizarre lands and finally ending in the Gödelian Islands giving a lucid explanation of Gödel's famous Incompleteness Theorem. The reasoning employed is always enlightening as Smullyan reaches for more general principles; for example, on page 110, it is demonstrated that if a speaker says, "If I am a Knight, then P," where P is any proposition, then the speaker must be a Knight and P must be true! This fact enables one to more easily solve several of the puzzles. Moreover, it is a simple example of "self-referential reasoning," where the speaker makes a statement that references himself. In this paper, we shall apply this kind of reasoning to identify faulty computers. We consider computers to be "faulty" if they always give the wrong aswer (Knaves) or sometimes give the wrong answer (Normals).

We wish to determine which processors (components, chips, etc.) are reliable in a remote location using as few "Yes or No" questions as possible. It is assumed that the processors have the ability to test each other. The questions will be asked by us, but answered by the processors, themselves, about each other. It may be that it is too dangerous for us to visit this remote location, in person, to test the processors; for example, we might need to test the reliability of a system in outer space or inside a nuclear reactor. It is further assumed that the majority of procesors are reliable. "Reliable" is often taken to mean "tells the truth;" however, in this paper, we shall take it to mean we are able to obtain correct information from it. We will explain this distinction below. This problem of determining which components are reliable by questioning the components themselves about each other has a long history (see, for example, [2],[3],[4],[5],[11]).

In this paper, the processors are assumed to be of three types which we term, a là Smullyan[5]: Knights, Knaves and Normals. "Knights," always tell the truth, "Knaves," always lie, and "Normals," sometimes tell the truth and sometimes lie. Using self-referential reasoning, we shall show how to find all the reliable ones, by asking only "Yes or No" questions, assuming that the Knights and Knaves costitute a majority, that is, the Normals make up less than half of the total number of processors. In fact, by using appropriately constructed self-referential questions, both the Knights and the Knaves will give us the same valid answers, that is, both the Knights and Knaves can be considered to be "reliable." This contrasts with the usual approach, where the processors are asked directly, "Is that processor a Knight?" and it is assumed that the Knights constitute a majority of the processors (see [2], [3],[11]).

Another way to look at the difference between our self-referential reasoning approach and the standard one is the following. In our approach, the interpretation of the word "reliable" is expanded. Instead of regarding just the Knights to be reliable and Knaves and Normals, unreliable, we consider both the Knights and Knaves to be reliable and only the Normals, to be unreliable. Thus, the sufficient condition, "the number of unreliables must be less than half the total number of processors," is more easily satisfied. For example, if we know there are 3 processors of each type, it is necessary to use self-referential questioning to find a reliable individual.

Our first algorithm, presented in the form of a "Smullyanesque" puzzle, finds all of the Knights. The second and third algorithms we present are modifications of standard algorithms using non-self referential reasoning. The second algorithm is more complicated than the first, but uses fewer questions, in the worst case. Finally we present an algorithm that eliminates just the Normals, allowing us to then question a reliable indi-vidual, that is, a Knight or a Knave.

Our self-referential questioning technique depends on a general principle, the "Nelson Goodman Principle," that elicits valid information from either a Knight or a Knave and we discuss it in the next section.

2. Nelson Goodman Principle(NGP).

Suppose we meet an individual who is either a Knight or a Knave, and we want to find out if statement P is true or false, by asking only one "Yes or No" question. Raymond Smullyan has pointed out (see [8],[9]) that it is indeed possible to do this and he attributes this observation to the logician and philosopher, Nelson Goodman and calls it the Nelson Goodman Principle. (We shall use the abreviation, NGP, for this principle.) Here is the question to be asked:

    (Q)    "Is it true that P if and only if you are a Knight?"

It is easy to see that a "Yes" answer to question Q from either a Knight or a Knave means that P is true; for, if a Knight answers "Yes," Q must be true; also, since "you are a Knight" is true, P must be true as well. If a Knave answers "Yes," Q must be false; since "you are a Knight" is false, P cannot be false, since then the biconditional would be true! Hence P is true in this case as well.

On the other hand, a "No" answer to Q from a Knight or Knave means that P is false; for, if a Knight answers "No," Q must be false and since "you are a Knight" is true, P must be false. If a Knave answers "No," Q must be true, and since "you are a Knight" is false, P must also be false.

3. Finding the Knights.

In the spirit of Raymond Smullyan's puzzle books, we present our first algorithm in the form of a puzzle.

The Puzzle. A heinous crime has been committed on a remote island and a Detective has been dispached from the mainland to investigate. Now, it happens that there are three types of inhabitants on this island: Knights, who always tell the truth, Knaves, who always lie, and Nor-mals, who sometimes tell the truth and sometimes lie. Also, it is known that fewer than half of the inhabitants on the ilsand are Normal. The Detective's immediate goal is to find the Knights, since they will answer her questions truthfully. It is impossible to distinguish the inhabitants by their appearance, but they are quite willing to answer "Yes or No" questions. Can she find all of the Knights?

Solution. Yes, it is possible to find all the Knights by just asking "Yes or No" questions. The Detective assembles all the inhabitants of the Island and she lines them up in a single line. She will first use the following question: "Is it true that the person who is next in line is a Normal if and only if you are a Knight?" A "Yes" answer to this question implies that either the respondent or the person next in line is a Normal; for, this is certainly true if the respondent is Normal; in the case that the respondent is a Knight or a Knave, the NGP implies that the next in line really is a Normal.

The Detective asks her question to the first in line. If the first person answers "No," she asks her question to the next person in line, and so on, until she receives a "Yes" answer or she has questioned everyone, except the last person in line. If she receives a "Yes," she removes both the respondent and the following person from the line. She then proceeds to question the person who was immediately before the pair that was removed, or, if there isn't such person, she questions the person who is now first in line. (Note that the person she is now going to question will answer about a "new" next in line. ) This mode of questioning continues until everyone remaining in line, except the last, has been questioned (and has answered "No"). Since the Normals constitue fewer than half the population of the island, and since at least half of those removed from the line are Normals, it cannot be the case that every Knight and Knave were removed from the line. Furthermore, the line cannot contain a Normal preceded by either a Knight or a Knave, since, in this case, the Knight or Knave would have answered "Yes" to the Detective's question and both would have been removed. This implies that the last in line is now either a Knight or a Knave, since otherwise the line would end with a string of Normals, preceded by a Knight or Knave which is impossible as we have seen.

Once the Detective can identify a Knight or Knave she can find all the Knights by asking the Knight or Knave the question: "Is it true that this person is a Knight if and only if you are a Knight?" The answer will be "Yes" if and only if the individual being discussed is a Knight. In this way all the Knights were identified and from the information she obtained from the Knights she was able to deduce the identity of the criminal who committed the "heinous crime" of making up this puzzle.

If there are n inhabitants of the island, the Detective has to ask n - 1 "Yes or No" questions before reaching the end of the line, since a "No" answer moves her one step closer to the end, while a "Yes" answer, moves her one step back, but removes two people further on, resulting in her again being one step closer to the end. Then all the Knights can easily be found with n additional questions, addressed to the Knight or Knave, about the n inhabitants of the island including itself. Thus (2n - 1) questions suffice to locate all the Knights.

We note that if we only wish to find one reliable inhabitant, where "reliable" means "Knight or Knave," then n - 1 questions suffice. In fact, it is possible to do a little better. We show next how to find a reliable inhabitant using n - h(n) questions, where h(n) is the number of 1s in the binary representation of n. (If n = 100, this results in a savings of two questions.) The algorithm we present uses a technique that has also been used to find a majority element in a set of differentiated elements (for example, colored balls [6]). We have adapted a version due to Steven Taschuk [10].

3. Finding a Knight or Knave with even fewer questions.

Again we assume that fewer than half the inhabitants of the island are Normal. This time our Detective proceeds as follows. (1) She arbitrarily groups the inhabitants of the island in (ordered) pairs, with possibly one left over. (2) For each pair she asks the first person the following question: "Is it true that the second member of this pair is reliable if and only if you are a Knight?" (Again, "reliable" means either a Knight or a Knave). By the Nelson Goodman Principle, a "Yes" answer, means that the second member is reliable, while a "No" answer means that the second member is Normal. If the answer is "No," she removes both members of the pair from further consideration. If the answer is "Yes," she removes just the first person. (3) If there was an unpaired person, she either keeps or removes this person, so as to maintain an odd number of people. We claim that after completing these three steps, it is still the case that fewer than half of the (remaining) inhabitants are Normal.

<u>Proof of Claim</u>. Let r, $r_2$, $r_3$ be the numbers of reliable inhabitants initially and after steps (2), (3), respectively. Let n, $n_2$, $n_3$ be the numbers of Normals, initially and after steps (2), (3), respectively. Then $r > n$ and we wish to show that $r_3 > n_3$.

Let #(R,R) be the number of ordered pairs produced by step (1) with both members reliable; let #(N,N) be the number of pairs with both members Normal; let #(R,N) be the number of ordered pairs whose first member is reliable and whose second member is Normal; let #(N,R) be the number of ordered pairs whose first member is Normal and whose second member is reliable.

In step (2), every (R,R) pair answers "Yes," so the second reliable inhabitant remains; this implies $r_2 \geq$ #(R,R). However, the only way for a Normal to survive step (2) is to be the second element of a (N,N) pair with the first Normal answering "Yes." This implies that $n_2 \leq$ #(N,N). There are three cases to be considered.

Case 1. There were an even number of inhabitants to be paired. Then r = 2 #(R,R) + #(R,N) + #(N,R) and n = 2 #(N,N) + #(R,N) + #(N,R). Since $r > n$, #(R,R) > #(N,N). Thus, $r_2 \geq$ #(R,R) > #(N,N) $\geq n_2$, that is, $r_2 > n_2$. There is no unpaired inhabitant, so $r_3 > n_3$.

Case 2. There were an odd number of inhabitants to be paired and the unpaired inhabitant was reliable. Then, r = 2 #(R,R) + #(R,N) + #(N,R) + 1 and n = 2 #(N,N) + #(R,N) + #(N,R). Since $r > n$, #(R,R) $\geq$ #(N,N) and this implies $r_2 \geq n_2$. If $r_2 > n_2$, then $r_3 > n_3$, whether or not we keep the unpaired reliable inhabitant. If $r_2 = n_2$, then the total number of inhabitants kept after step (2) is even and so we keep the unpaired reliable inhabitant and $r_3 = r_2 + 1 > n_2 = n_3$.

Case 3. There were an odd number of inhabitants to be paired but the unpaired inhabitant was Normal. Then, r = 2 #(R,R) + #(R,N) +

#(N,R)  and n = 2 #(N,N) + #(R,N) + #(N,R) +1. Therefore, since r > n, 2 #(R,R) > 2 #(N,N) + 1. So #(R,R) >  #(N,N). Thus,  $r_2 \geq$ #(R,R) > #(N,N)  $\geq n_2$, that is, $r_2 > n_2$. If  $r_2 = n_2 + 1$, then the total number of inhabitants after step (2) is odd and so the unpaired Normal is removed. If  $r_2 > n_2 + 1$, then $r_3 > n_3$, in any case.

Each time this three step procedure is repeated, at least half of the inhabitants are removed from consideration. Finally, the Detective is left questioning the sole remaining inhabitant, who must be reliable.

We claim that, in the worst case, when in step (2) only one person is removed from each pair because all questioned by the Detective answer "Yes," the number of questions is n - h(n), where n is the number of inhabitants on the island and h(n) is the number of 1s in the binary representation of n. Assume this is true for all k < n.  If n is even, step (1) results in n/2 pairs with no one left over, n/2 questions having already been asked by the Detective and n/2 inhabitants remain (assuming all "Yes" answers). Since n/2 < n, our induction assumption implies that n/2 - h(n/2) further questions are needed in the worst case to find a reliable person. However the binary representation of n/2 has same number of 1s as the binary representation of n. Thus the total number of questions asked is at most n/2 + n/2 - h(n/2) = n - h(n). If n is odd, step (1) results in $\lfloor n/2 \rfloor$(the floor of n/2) pairs and h($\lfloor n/2 \rfloor$) = h(n) - 1. Thus the total number of questions, when n is odd, is $\lfloor n/2 \rfloor$ + $\lfloor n/2 \rfloor$ - h($\lfloor n/2 \rfloor$) = n - 2 - h(n) + 1 < n - h(n). Thus, in the worst case, n - h(n) questions suffice.

4. Eliminating the Normals.

 We now turn to the problem of eliminating just the Normals. Our interrogation strategy is an adaptation of that of  Blecher [3] (rediscovered by Wildon [11]) .

To facilitate finding the Normals, we make use of the NGP asking the following self-referential question, question Q.

   Q:   Is it true that  processor X is Normal if and only if you are a Knight?

The NGP implies that if either a Knight or Knave answers "Yes" to Q,

then processor X is indeed Normal; if a Knight or Knave answers "No," processor X must not be Normal, that is, processor X is a Knight or a Knave.

Theorem. Suppose that there are a total of n processors, u of which are Normal, where u < n/2. Then fewer than $\frac{3}{2}$n "Yes or No" questions suffice to identify all the Normal processors.

Proof. Assume that the Theorem is true for all sets of processors of size less than n and let B be a set of processors of size n.

Choose a processor X, whose status is unknown and ask the other processors, in turn, question Q about processor X. We stop as soon as either of the following two conditions is satisfied.
(a) u processors have answered "No."
(b) More processors have answered "Yes" than have answered "No."
(One of these two cases must occur, since u < n/2; for, if fewer than u processors answer "No," then more than half the processors answer "Yes" and so, at some point, more processors will have answered "Yes" than "No.")

Suppose we stop in case (a). Then the processor X must be either a Knight or a Knave; for if X were Normal, at least one of those answering "No" to Q must not be Normal, since there are only u Normals including X; hence one of these respondents must be either a Knight or a Knave; but then, if a Knight or Knave answers "No," X cannot be Normal, by the NGP. Note that, in this case, if u + a questions have been asked, with a having anwered "Yes," these a who answered "Yes" must be Normals. We then ask processor X about the other (n - a) processors, using question R.

R: Is it true that processor Y is a Knight or a Knave if and only if you are a Knight?

A "Yes" answer to R will mean that Y is a Knight or a Knave; a "No" answer implies Y is Normal. In this way we have found all the Normals using (n - a) + (u + a) = n + u < $\frac{3}{2}$n questions.

Suppose we stop in case (b) and C is the set of those processors in B who have answered "Yes" to question Q and D is the set of those in B

who have answered "No" to Q; thus $|C| = |D| + 1$. Assume first that at least one in C is a Knight or Knave; then, the "Yes" answer to Q from this processor implies that X is Normal (by NGP); moreover, all those in D, having answered "No," can't be either a Knight or a Knave (NGP) and hence also must be Normal. Thus, in this case, at least $|D| + 1$ of the processors in $C \cup D \cup \{X\}$ are Normal, that is, at least half of the processors in $C \cup D \cup \{X\}$ are Normal. If, however, no one in C is a Knight or Knave, then C consists entirely of Normals and again, at least half of the processors in $C \cup D \cup \{X\}$ are Normal. Therefore, if the processors in $C \cup D \cup \{X\}$ are removed from B, the reduced set of processors, $E = B - (C \cup D \cup \{X\})$, also obeys the condition $u_1 < n_1/2$, where $n_1$ is the number of processors in E and $u_1$ is the number of Normals in E. Note that $n_1 < n$. By our induction assumption, fewer than $\frac{3}{2} n_1$ questions suffice to find the all Normals in E. It remains to find the Normals in $C \cup D \cup \{X\}$. Since $u_1 < n_1/2$, E must contain non Normal processors. We then choose a non Normal processor, processor Z, from E and ask question Q to processor Z about processor X. This allows us to determine X's status and uses one additional question. There are two cases.

1) If X is Normal, those in set D, that is, those who said "No" to Q, can't be Knights or Knaves, and hence must be Normals. We then ask processor Z question Q about each processor in set C. This uses $|C|$ additional questions.

2) If X is a Knight or a Knave, those in set C, having said "Yes" to Q must be Normals. In this case, we ask processor Z question Q about each processor in D. This uses $|D|$ additional questions.

Since $|B| = n$ and $|E| = n_1$, $|C \cup D \cup \{X\}| = n - n_1$. But, $|C| = |D| + 1$. So, $|D| < \left(\frac{n - n_1}{2}\right)$ and $|C| \leq \left(\frac{n - n_1}{2}\right)$.

Therefore, in either case 1) or case 2), we require, at most, an additional $\left(\frac{n - n_1}{2}\right)$ questions. Since these cases are mutually exclusive, we require fewer than $\frac{3}{2} n_1 + 1 + \left(\frac{n - n_1}{2}\right) = n_1 + \frac{n}{2} + 1$ questions. But $n_1 < n$, so $n_1 + \frac{n}{2} + 1 < \frac{3}{2} n + 1 \leq \frac{3}{2} n$. Thus we require fewer than $\frac{3}{2} n$ questions, as claimed.

* Professor Emeritus
Queens College, CUNY
email: robert.cowen@gmail.com